\documentstyle[12pt,preprint]{aastex}
\input{epsf.sty}

\def\kms{km\thinspace s$^{-1}$}     
\def\deg{\ifmmode^\circ\else$^\circ$\fi}  
\def\arcs{\ifmmode {'' }\else $'' $\fi}  
\def\arcm{\ifmmode {' }\else $' $\fi}    

\def\lya{Ly$\alpha$}
\def\lyb{Ly$\beta$}
\def\cm2{cm$^{-2}$}
\def\NHI{$N_{\rm HI}$}
\def\lNHI{log~$N_{\rm HI}$}

\def\etal{et al.\ }

\begin{document}

\title{The Baryon Content of the Local Intergalactic Medium}
\author{John T. Stocke, J. Michael Shull, \& Steven V. Penton} 
\affil{Center for Astrophysics \& Space Astronomy, and Dept. of 
Astrophysical \& Planetary Sciences, University of Colorado,
Boulder, CO 80309-0389}

\begin{abstract}

In this review, we describe our surveys of low column density 
(\lya) absorbers (\NHI $= 10^{12.5-16}$~cm$^{-2}$), which show that 
the warm photoionized IGM contains $\sim$30\% of all baryons at 
$z \leq 0.1$.  This fraction is consistent with cosmological
hydrodynamical simulations, which also predict that an additional 
20--40\% of the baryons reside in much hotter 10$^{5-7}$~K gas, 
the warm-hot IGM (WHIM).  The observed line density of \lya\ absorbers, 
$d{\cal N}/dz \approx 170$ for \NHI $\geq 10^{12.8}$ \cm2, is dominated
by low-\NHI\ systems that exhibit slower redshift evolution than 
those with \NHI $\geq 10^{14}$~\cm2.  HST/FUSE surveys of O~VI absorbers, 
together with recent detections of O~VII with {\it Chandra} and
{\it XMM/Newton}, suggest that anywhere from 20--70\% (with large
errors) of the baryons could reside in the WHIM, for an assumed 
abundance O/H $\approx$ 10\% solar.  
We also review the relationship between the various types of \lya\ 
absorbers and galaxies. At the highest column densities, 
\NHI\ $\geq 10^{20.3}$~cm$^{-2}$, the damped \lya\ (DLA) systems 
are often identified with gas-rich disks of galaxies over a large 
range in luminosities (0.03--1~$L^*$) and morphologies.  Lyman-limit systems 
(\NHI $= 10^{17.3-20.3}$~cm$^{-2}$) appear to be associated with bound 
bright ($\geq$ 0.1--0.3 $L^*$) galaxy halos.  The \lya\ absorbers with 
\NHI $=10^{13-17}$~cm$^{-2}$ are associated with filaments of
large-scale structure in the galaxy distribution, although some may arise 
in unbound winds from dwarf galaxies.  Our discovery that $\sim20$\% of 
low-$z$ \lya\ absorbers reside in galaxy voids suggests that a substantial 
fraction of baryons may be entirely unrelated to galaxies.  In the future, 
HST can play a crucial role in a precise accounting of the local baryons
and the distribution of heavy elements in the IGM.   These studies will
be especially effective if NASA finds a way to install the {\it Cosmic 
Origins Spectrograph} (COS) on {\it Hubble}, allowing an order-of-magnitude 
improvement in throughput and a comparable increase in our ability to study 
the IGM.

\end{abstract}


\section{Introduction}

In its first year, the {\it Hubble Space Telescope} (HST) discovered
that a majority of all baryons in the current universe are not in
galaxies, but instead remain in the intergalactic medium (IGM).
In subsequent years, the UV spectrographs aboard HST and the
{\it Far Ultraviolet Spectroscopic Explorer} (FUSE) have continued
these investigations of the multiphase IGM, using sensitive
UV tracers of diffuse gas: the Lyman series of H~I (\lya\ at 1215.67~\AA,
\lyb\ at 1025.72~\AA, etc) and the O~VI doublet (1031.926, 1037.617~\AA).
These HST and FUSE studies have led to a preliminary ``baryon census''
of the ``warm'' (photoionized) and ``warm-hot'' (collisionally ionized) IGM.  
With spectrographs aboard the {\it Chandra} and {\it XMM/Newton} X-ray 
telescopes, astronomers are beginning to search for even more highly 
ionized gas through resonance absorption lines of O~VII, O~VIII, N~VII,
and Ne~IX.

Unlike virtually all other astronomical objects, the Ly$\alpha$ absorption
systems were first discovered at great distances ($z \geq 2$) owing to
their cosmological redshifts and the near-UV atmospheric cutoff. Only 
with the advent of HST have nearby examples been found.  The 
first low-$z$ \lya\ absorbers were seen in the spectrum of 3C~273 at 
$z_{\rm abs} < 0.158$ (Bahcall et~al.\ 1991; Morris et~al.\ 1991).
While the number of absorbers was significantly less than the line density 
at high-$z$, the ``local \lya\ forest'' contains far more absorbers than 
expected from extrapolating the ground-based data (Bahcall \etal 1993 and 
subsequent QSO Absorption-Line Key Project papers by Jannuzi \etal 1998 
and Weymann \etal 1998).  Although the \lya\ absorbers at $z \geq 2$ are 
sufficiently abundant to account for nearly all the baryons (Rauch \etal 
1997; Schaye 2001), their 
substantial numbers at $z \leq 0.1$ imply that $\sim$30\% of all baryons 
remain in these photoionized clouds locally (Penton, Stocke, \& Shull 2000a,
Penton, Shull, \& Stocke 2000b, 2004, hereafter denoted Papers I, II, and IV).

Numerical simulations (Fig.\ 1) of the evolving IGM (Cen \& Ostriker 1999; 
Dav\'e et~al. 1999, 2001) explain not only the general features of the \lya\ 
number density evolution, but also many detailed properties, including distributions
in column density (\NHI) and doppler $b$-value (Papers II and IV; 
Dav\'e \& Tripp 2001; Ricotti, Gnedin, \& Shull 2000), and their relationship 
to galaxies (Dav\'e \etal 1999; Impey, Petry, \& Flint 1999; 
Penton, Stocke, \& Shull 2002, hereafter
denoted Paper III). Any accounting of the present-day distribution of baryons 
must include an accurate census of these absorbers and the associated mass, 
inferred from their ionized fractions, column densities, and physical extents.

\begin{figure}[t] \epsscale{0.6}
\plotone{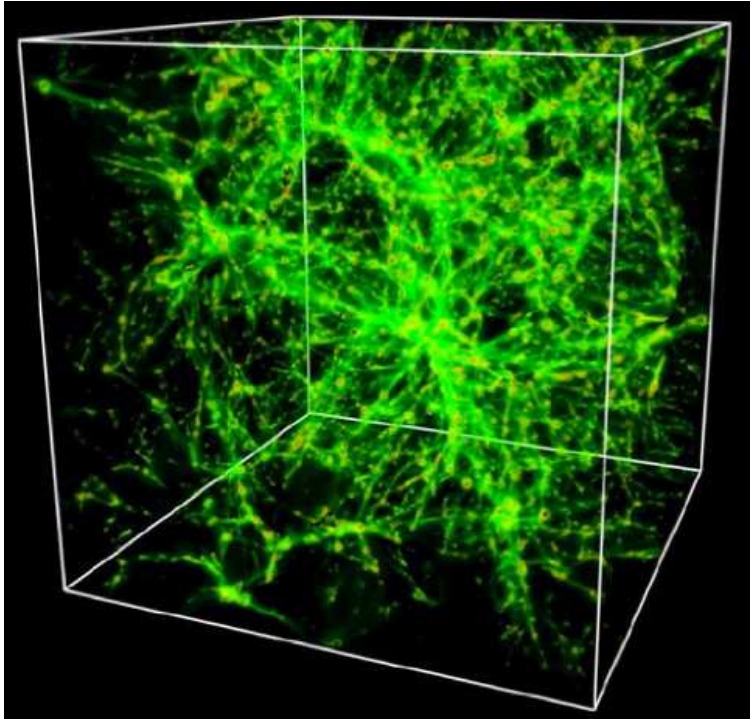}
\caption{\small The distribution of baryons in the ``cosmic web" of dark
 matter (Cen \& Ostriker 1999).  In this simulation the Ly$\alpha$,
 O~VI, and other absorption lines are produced in the denser filaments, 
 representing a fluctuating distribution of gas structured by the dark-matter 
 gravitational potentials. }
\end{figure}

Moderate-resolution UV spectroscopy of bright quasars, QSOs, blazars,
and Seyfert galaxies has provided a substantial database of low-$z$ \lya\ 
absorbers.  At the same time, several ground-based galaxy surveys (Morris 
\etal 1993; Lanzetta \etal 1995; Chen \etal 1998; Tripp, Lu, \& Savage 
1998; Rao \& Turnshek 1998, 2000; Impey, Petry, \& Flint 1999;  
Nestor \etal 2002; Bowen \etal 2002; Paper III; Bowen \& Pettini 2003;
Stocke \etal 2005, hereafter denoted Paper~V) have probed the relationship 
between \lya\ absorbers and galaxies, filaments of galaxies, and voids. 
Using nearby examples of the \lya\ phenomenon, these authors sought to identify 
the galaxy types responsible for the absorption and thus assist in interpreting 
the wealth of information (number densities, metallicities, ionization states,
line widths) of \lya\ absorption systems at high-$z$. These efforts have been
somewhat successful, although the results in most column-density regimes remain 
controversial (see conference proceedings edited by Mulchaey \& Stocke 2002). 

In this review, we describe the various HST QSO absorption line surveys that 
have been undertaken (\S~2), review our present knowledge of the baryon content 
of the IGM (\S~3), and describe the emerging, but still controversial, evidence 
for the relationship between the various column densities of \lya\ absorbers and 
galaxies (\S~4). The last information has come largely from studying low-$z$ 
absorbers discovered with HST. We conclude (\S~5) 
with a brief prospectus on low-$z$ 
IGM studies facilitated by the {\it Cosmic Origins Spectrograph} (COS), 
a new instrument that may be installed on HST in the coming years.

\section{HST Surveys the Low-$z$ \lya\ Absorbers}

The HST with its UV spectrographs (FOS, GHRS, STIS) conducted several important 
surveys of the IGM, which provided basic data for studying the bulk of local 
baryons. Owing to its modest spectral resolution (200-300 \kms), the {\it Faint 
Object Spectrograph} (FOS) used for the initial QSO Absorption-Line Key Project 
(Bahcall \etal 1993) primarily detected high column density \lya\ absorbers
with equivalent widths $W_{\lambda} \geq 240$~m\AA.  
The Key Project provided examples of the various types of \lya\ absorbers: 
damped \lya\ (DLA) absorbers, Lyman-limit/strong Mg~II absorbers, weak 
high-ionization (C~IV) and low-ionization (Mg~II) metal-line absorbers, and 
\lya-only absorbers (Bahcall \etal 1996; Jannuzi \etal 1998). Even though 
the broad UV wavelength coverage (G130H, G190H, G270H gratings) of the Key 
Project spectra allowed the discovery of many \lya\ absorbers at 
$z \leq 1.6$, the detection efficiency of low redshift ($z \leq 0.2$) 
absorbers was reduced by lower than expected far-UV sensitivity of the FOS 
digicon. The FOS Key Project survey firmly established the column density 
distribution, $f$(\NHI), for high-\NHI\ absorbers and $d{\cal N}/dz$, the 
number density of absorbers per unit redshift.  Above limiting \lya\ equivalent 
width, $W_{\lambda} = 240$~m\AA\ (\NHI $\geq 10^{14}$~cm$^{-2}$), Weymann \etal 
(1998) found $d{\cal N}/dz = 32.7 \pm 4.2$ over a substantial redshift 
pathlength ($\Delta z \approx$ 30).  As we discuss below,  
the \lya\ line density increases substantially to lower columns, reaching
$d{\cal N}/dz \approx 170$ for \NHI $\geq 10^{12.8}$~\cm2 (Paper~IV).    

The absorber number density (Weymann \etal 1998) shows a dramatic break 
from rapid evolution ($z \geq 1.5$) to almost no evolution ($z \leq 1.5$).  
This observation was explained by cosmological hydrodynamical simulations 
(Dav\'e \etal 1999), in which the rapid evolution in $d{\cal N}/dz$ is 
controlled by the response of photoionized gas to evolving QSO populations 
in a hierarchical distribution of large-scale structure.  At high redshift,
the number of \lya\ absorbers decreased rapidly with time, owing to a nearly 
constant extragalactic ionizing flux from QSOs in an expanding universe. 
[The cosmological expansion decreases recombinations, thus decreasing the 
neutral hydrogen column density.] At $z \leq 2$, the ionizing flux began 
a rapid decline with the rapidly decreasing QSO numbers and/or luminosities, 
thereby decreasing the ionized fraction in the \lya\ absorbers.  As a 
consequence of reduced photoionization, along with large-scale structure 
formation, the $d{\cal N}/dz$ evolution slowed rapidly.
Although the Key Project discovered only one DLA (Jannuzi \etal 1998), 
Rao \& Turnshek (1998) used FOS to increase the number of low-$z$ DLAs by 
targeting strong Mg~II absorbers found by ground-based telescopes. Bowen \etal 
(1996) used archival FOS spectra to search for strong \lya\ and metal-line 
absorption from known foreground galaxies. See Bowen, Pettini, \& Blades 
(2002) for a STIS extension of this successful project. 

The {\it Goddard High-Resolution Spectrograph} (GHRS) 
was used to co-discover (Morris \etal 1991) the \lya\ 
absorbers in the 3C~273 sightline.  However, few QSOs were sufficiently bright 
to observe in moderate-length integrations. Tripp, Lu, \& Savage (1998) and 
Impey, Petry, \& Flint (1999) observed a few UV-bright AGN targets with the 
low-resolution far-UV GHRS grating, obtaining observations only slightly more 
sensitive than the FOS Key Project spectra. In the mid-1990s, the Colorado 
group conducted a moderate-resolution ($\sim$ 20 \kms) survey of 15 sightlines 
with the GHRS/G160M first-order grating (Papers I and II), followed by a 
comparable survey of 15 additional targets with the STIS first-order, 
medium-resolution grating G140M (Paper IV). In these surveys, the key  
strategy was to select only the brightest targets (see Fig.\ 2), while 
observing only a modest pathlength per target ($cz \approx$ 10,000 \kms\ 
with GHRS and 20,000 \kms\ with STIS). The number density of \lya\ 
absorbers rises steeply with decreasing equivalent width.  After subtracting 
intrinsic absorbers (near the redshift of the QSO) we found 187 intervening
low-\NHI\ \lya\ absorbers at $\geq 4\sigma$ significance over a total unobscured 
pathlength $\Delta z =1.157$.  A careful analysis, accounting for S/N sensitivity
bias and overlap with Galactic interstellar lines, gives a line density
$d{\cal N}/dz \approx 170$ for \NHI $\geq 10^{12.8}$~cm$^{-2}$, 
or one absorber every $\sim 25~h^{-1}_{70}$ Mpc.  Our more sensitive 
GHRS/STIS survey provides an important extension of the Key Project 
to lower-\NHI\ absorbers. As shown in Figure 2, the very best spectra in 
our survey reach equivalent-width limits of $W_{\lambda} =$ 10-20~m\AA\ 
($4 \sigma$), corresponding to \NHI $\approx 10^{12.3-12.5}$ \cm2, 
comparable to the best spectra obtained with Keck/HIRES or VLT/UVES.

\begin{figure}[t] 
\epsscale{0.6}
\plotone{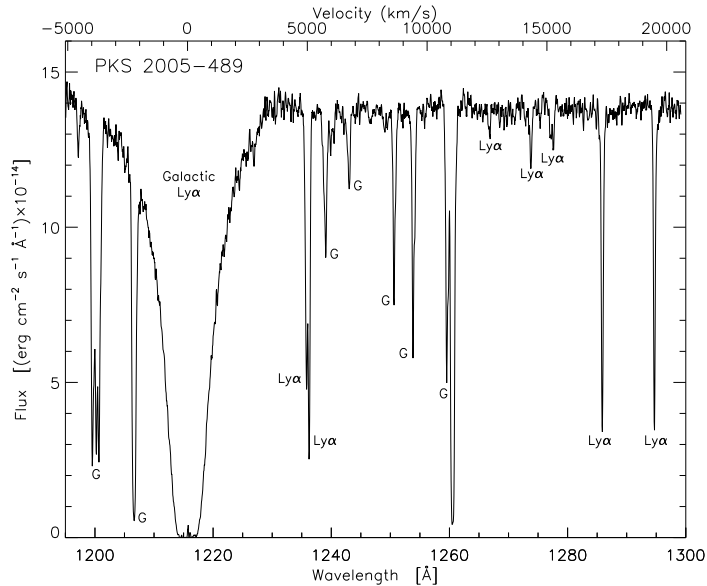}
\caption{\small
An HST/STIS medium-resolution (20 \kms) spectrum (Paper IV)
of the bright BL Lac Object PKS 2005-489 illustrates the highest-quality 
data obtained for this project. The deep, broad absorption
at 1216 \AA\ is the damped Ly$\alpha$ absorption due to the Milky Way. 
Other Galactic interstellar metal lines (S~II, Si~II, Mg~II,
N~I, N~V and Si~III) are marked with a ``G''. The weakest Ly$\alpha$
absorbers are at 1266.740~\AA\ (12,594 \kms, $W_{\lambda} = 22 \pm 13$~m\AA)   
and 1277.572~\AA\ (15,265 \kms, $W_{\lambda} = 27 \pm 12$~m\AA) with
column densities \NHI $=10^{12.6-12.7}$~\cm2). The heliocentric velocity 
scale along the top is for the Ly$\alpha$ absorbers only.}
\end{figure}
 
The cosmic evolution of the Ly$\alpha$\ forest absorbers shown in Figure 3 
represents the historical record of a large fraction of the baryons during the
emergence of the modern universe at $z \leq 2$. The $d{\cal N}/dz$ evolution 
of the high-\NHI\ absorbers exhibits a substantial break in slope to a much 
slower evolution at $z\leq$1.5, as expected from the physical effects acting 
on intergalactic absorbers in an expanding universe.
The low-\NHI\ data point at $z \approx 0$ from our survey is lower than that
of the Key Project. We suspect this results from our survey's higher spectral 
resolution, which affects the counting and \NHI-bin assignments of weak 
absorbers.  The break in the $d{\cal N}/dz$ evolution of the higher-\NHI\ 
absorbers is therefore not as dramatic as the Key Project 
data suggest. A more gradual decline in slope is also more consistent with 
the simulations in a $\Lambda$CDM Universe model (Dav\'e et~al. 1999). 
The physical effects that account for the break in the high-column 
$d{\cal N}/dz$ should create a similar break in the evolution of the 
lower-\NHI\ absorbers, although this is not yet evident in Figure 3. 
New STIS spectra obtained by B. Jannuzi and collaborators should determine 
whether the expected break is present, and whether the numerical simulations 
have used the correct physics to account for these absorbers.

\begin{figure}[t] 
\epsscale{0.6}
\plotone{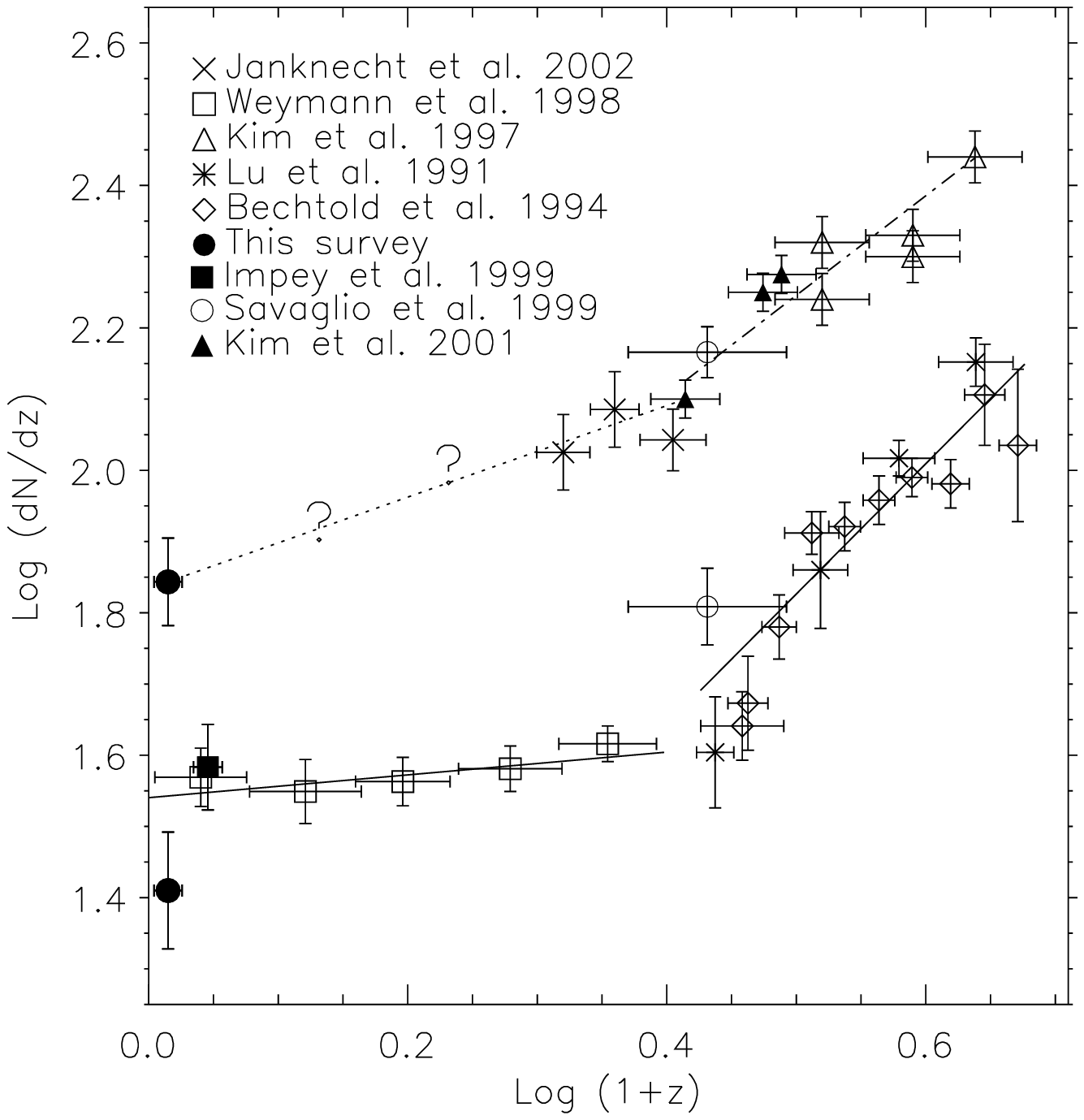}
\caption{\small
Comparison of the \lya\ absorber evolutionary plot, $dN/dz$ versus log(1$+z$),
for two N$_{H I}$ ranges.  The lower distribution corresponds to 
W$_{\lambda}\geq$240~m\AA\ (\NHI $\geq 10^{14}$~\cm2 for an assumed 
$b =25$ \kms).  The upper distribution corresponds to absorbers in the range 
\NHI $=10^{13.1-14.0}$~\cm2.  The two $z \approx 0$ points 
({\bf solid circles}) are taken 
from our survey (Paper IV) for each of the two \NHI\ ranges specified above.
Solid lines are taken from Weymann et al. (1998) and have evolutionary slopes
[$d{\cal N}/dz \propto (1+z)^{\gamma}$] with $\gamma = 0.16$ at $\log (1+z)< 0.4$ 
and $\gamma = 1.85$ at $\log (1+z)> 0.4$.  The complete evolutionary picture for 
low-\NHI\ absorbers is unavailable, owing to the lack of data at intermediate 
redshifts, as indicated by question marks (see Paper IV for details).}
\end{figure}

Recently, several investigators have used STIS in its medium-resolution 
($\sim 7$ \kms) echelle modes (E140M and E230M) to obtain long integrations on 
bright AGN targets.  Many of these spectra have been analyzed for 
interesting individual absorbers, including a growing number of redshifted
O~VI absorbers : H1821+643 (Tripp \etal 2000, 2001); 3C~273 (Sembach \etal 
2001; Tripp \etal 2002); PG~0953+415 (Savage \etal 2002); 
PKS~2155-304 (Shull \etal 2003); PG~1259+593 (Richter \etal 2004);
PG~1116+215 (Sembach \etal 2004); and PG~1211+143 (Tumlinson \etal 2004).  
These spectra have the potential 
to be HST's best resource for low-$z$ \lya\ absorber studies. As of July 
2004, 35 bright AGN have been observed in STIS-echelle mode, providing 
in some cases the most sensitive detection limits for \lya, as well as 
significant pathlengths at $z \geq 0.11$ for potential discovery of O~VI 
absorbers. In the absence of the {\it Cosmic Origins Spectrograph}, the 
STIS medium-resolution 
echelle gratings continue to provide high-quality QSO spectra for IGM studies.
However, obtaining the S/N needed to make sound inferences about the
IGM absorbers requires increasingly long integration times on fainter targets.   

\section{Baryon Content of the low-$z$ Ly$\alpha$ Forest}

At high redshift ($z =2 - 4$) the number density of \lya\ absorbers is
so large that estimates of their total baryonic masses, after substantial
ionization corrections for photoionization equilibrium, suggest that the 
entire baryon content of the universe ($\Omega_b h_{70}^2 = 0.045 \pm 0.003$, 
as measured by the Cosmic Microwave Background anisotropies or cosmic 
[D/H] ratio) can be accomodated within the warm ($\sim 10^4$~K) photoionized 
IGM (Rauch \etal 1997; Schaye 2001). This result is consistent with numerical 
simulations of large-scale structure formation (Cen \& Ostriker 1999; 
Dav\'e \etal 1999, 2001), which also predict that by $z \approx 0$ the baryons
are approximately evenly divided among several components as follows:  
\begin{itemize}

\item {\bf Stars and gas in or near galaxies; 30\% predicted}. 
Salucci \& Persic (1999) and Fukugita (2004) account for only $\sim 6$\% 
in this reservoir. 

\item {\bf Very hot ($10^{7-8}$~K) intracluster and intragroup gas;
$\leq$10\% predicted}. Fukugita (2004) estimates only $\sim5$\% 
of the baryons in this component, but the gas in groups may not be 
fully accounted for. 

\item {\bf Warm ($10^4$~K) photoionized gas; 30--40\% predicted in 
\lya\ forest absorbers}. In our Paper IV, we find $29 \pm 4$\% of the
baryons in local \lya\ absorbers. 

\item {\bf Warm-hot intergalactic medium (WHIM) at $10^{5-7}$~K;
20--40\% predicted}.  A series of observations with HST and FUSE 
(Tripp, Savage, \& Jenkins 2000; Savage \etal 2002) estimate that 5-10\% 
of the baryons have been detected (at 10\% assumed O/H metallicity) 
through O~VI absorption. Recent {\it Chandra} 
observations of O~VII absorption (Fang \etal 2002; Nicastro \etal 
2004) suggest even larger percentages, although with enormous statistical 
uncertainties and a few unconfirmed observations. 

\end{itemize}
Thus, we submit that the current-epoch baryon census remains an unsolved 
problem, with large and uncertain corrections for ionization state 
(H~I, O~VI, and O~VII) and an overall uncertainty over the
appropriate metallicity (for O~VI, O~VII, O~VIII in WHIM).
These problems can be addressed by HST and FUSE in the coming 
years, and by future high-throughput X-ray spectroscopic missions such as
{\it Constellation-X} or {\it Xeus}. 

\subsection{Photoionized \lya\ Absorbers}

In Paper II, we described a method for estimating the baryon content of the 
local \lya\ absorbers, based upon the observed column density distribution. 
In the context of optically-thin, photoionized clouds, the assumptions of this 
simplified model are: (1) spherically symmetric absorbers; (2) an isothermal 
density profile; (3) absorber sizes of $100 h_{70}^{-1}$~kpc, based on QSO 
pairs experiments conducted at somewhat higher redshifts (Dinshaw \etal 1997, 
but see also Rosenberg \etal 2003); and (4) a value and slope for the
extragalactic ionizing flux (Shull \etal 1999) using space densities and
ionizing spectra of local Seyfert galaxies (Telfer \etal 2002).  However,
numerical simulations show that \lya\ absorbers are not at all spherical, 
especially at lower column densities. Cen \& Simcoe (1997) used 
numerical simulations at $z$=2--4 to show that lower column density 
(\NHI $= 10^{13-14}$~\cm2) absorbers have large physical sizes, consistent 
with or somewhat larger than $100 h_{70}^{-1}$~kpc (equivalent spherical radius).  
We have integrated this simplified model over our observed column density 
distribution, from 12.5 $\leq$ log~\NHI $\leq 16$, the range where the above 
assumptions are most valid.  In Paper~II we obtained a baryon fraction 
$\geq20$\%, with specific dependences on measurable parameters as follows:
\begin{equation}
   \Omega_{{\rm Ly}\alpha} = (0.008 \pm 0.001) \left[N_{14} \; I_{-23} \; b_{100} 
     \left( \frac {4.8} {\alpha_s + 3} \right) \right]^{1/2} h_{70}^{-1} \; .  
\end{equation}
Here, $N_{14}$ is the characteristic H~I column density in the \lya\ forest
in units of $10^{14}$~\cm2, $I_{-23}$ is the extragalactic ionizing specific
intensity at 1 ryd in units of $10^{-23}$ erg~cm$^{-2}$ s$^{-1}$ Hz$^{-1}$ 
sr$^{-1}$, $b_{100}$ is the characteristic absorber radius in units of 
$100h_{70}^{-1}$ kpc, and $\alpha_s \approx 1.8$ is the mean power-law spectral 
index ($F_{\nu} \propto \nu^{-\alpha_s}$) for the metagalactic ionizing 
radiation, which presumably comes from QSOs (Shull \etal 1999; Telfer \etal 2002).

Schaye (2001) developed a methodology for estimating the baryon content that
makes somewhat different assumptions. As with the method of our Paper II, this 
model assumes photoionization of optically-thin absorbers, but it also assumes 
gravitationally-bound clouds whose observed column densities are equal to their 
characteristic column densities over a Jeans length.  This estimate  
has the following specific dependences:
\begin{equation} 
  \Omega_{{\rm Ly}\alpha} = (2.2 \times 10^{-9}) h_{100}^{-1} \Gamma_{-12}^{1/3} 
     T_4^{0.59} \int {N_{\rm HI}}^{1/3} f(N_{\rm HI}, z) d N_{\rm HI} \; ,   
\end{equation}
where $\Gamma_{-12}$ is the H~I photoionization rate in units of 
$10^{-12}$~s$^{-1}$, $T_4$ is the IGM temperature in units of $10^4$~K,
and $f(N_{\rm HI}, z) \propto N_{\rm HI}^{-\beta}$ is the H~I column 
density distribution.  In this model, the absorber baryon content 
depends on the minimum column density of H~I absorbers as
$N_{\rm min}^{-(\beta - 4/3)}$ or $N_{\rm min}^{-0.32 \pm 0.07}$ for 
$\beta = -1.65 \pm 0.07$ (see below). The above 
expression assumes that the \lya\ absorbers contain the universal ratio 
of baryons to dark matter, with no bias.

Here we employ both of these methods to estimate the baryon content of the local 
\lya\ absorbers, based on our latest column density distribution from Paper IV, 
which fitted power-law slopes of $\beta = -1.65 \pm 0.07$ and $-1.33 \pm 0.30$ 
on the lower and upper sides, respectively,
of a break at \NHI $= 10^{14.5}$~\cm2. From our 
enlarged sample, we are confident that the $\beta = -1.65$ slope extends down 
to at least log~\NHI = 12.5, so that we can reliably integrate this distribution 
from log~\NHI = 12.5 to 16.0.  We recognize that, at or near our adopted lower 
limit (which corresponds approximately to an overdensity $\delta \approx 3$ 
at the current epoch), some of the assumptions of both methods may break 
down. Also, at \NHI $\geq 10^{17}$~\cm2, absorbers become 
optically thick in the Lyman continuum and can hide additional gas mass. 
Therefore, neither method can hope to derive an extremely accurate 
measurement for the baryon content of the \lya\ forest, even if the column 
density distribution and other absorber properties (size, ionizing flux, 
temperature) are known to some precision. 

We quote the resulting baryon densities as percentages of the total baryon 
density from measurements of the Cosmic Microwave Background anisotropy 
($\Omega_b h_{70}^2= 0.0457 \pm 0.0018$, Spergel \etal 2003).  A slightly 
smaller total baryon density has recently been obtained by measurement of 
D/H in a high-$z$ QSO absorber ($\Omega_b h_{70}^2= 0.0437 \pm 0.004$, 
Kirkman \etal 2003).  
The baryon fraction in the IGM is dominated by the lower column density 
absorbers (22\% out of 29\% total baryons in \lya\ absorbers at \NHI\ 
$\leq 10^{14.5}$~\cm2), owing to the steep slope of the column density 
distribution. The Schaye (2001) prescription yields a divergent baryon 
fraction at low \NHI, owing to an inverse dependence of size on column 
density, which must break down well above the limit employed here. On the 
other hand, at \NHI $\geq$ 10$^{14.5}$ cm$^{-2}$, the 100 kpc size assumed 
in Paper II probably becomes too large for many absorbers at higher column 
densities (Cen \& Simcoe 1997; Tripp \etal 2002). Thus, the most
conservative, and we believe most accurate approach, is to use our 
Paper II method at low columns and the Schaye (2001) method at high columns.
This technique yields values of 22$\pm$2\% below \NHI $= 10^{14.5}$~\cm2, 
and $7\pm3$\% above that column density, for a total baryon fraction in the 
low-$z$ photoionized IGM of $29\pm4$\% (Paper IV). Because the baryon fraction 
is dominated by the lowest column density absorbers, our result is relatively 
insensitive to the assumed $b$-values for individual absorbers. Here, we have 
used $b = 25$ \kms, which is comparable to the median $b =22$ \kms\ found by 
Dav\'e \& Tripp (2001) using medium-resolution echelle spectra.

In order to make further progress in the baryon census of warm 
IGM gas, it is important to: (1) estimate absorber sizes and shapes 
(and thus equivalent cloud radii) as a function of column density. In the 
absence of observations of many QSO pairs, this is best done using numerical 
simulations (the Cen \& Simcoe [1997] analysis, but for simulations at 
$z \approx 0$); 
(2) estimate the percentage of \lya\ absorbers that are collisionally
ionized rather than photoionized. This will require observations of 
O~VI with FUSE or HST to address the ``double-counting'' issue; see \S~3.2; 
(3) observe a much longer pathlength for \lya\ absorbers, since we find that 
the bias introduced by cosmic variance is still significant for the observed
redshift pathlength, $\Delta z = 1.157$ (Paper IV); (4) determine more
accurate H~I column densities from FUSE observations of higher Lyman lines 
and a curve-of-growth analysis (e.g., Shull \etal 2000; Sembach \etal 2001) 
for saturated absorbers; and (5) observe a few very bright targets over very 
long integration times ($\geq 30$ orbits) to push the detection limits below 
\NHI $\leq 10^{12}$ \cm2; very weak absorbers can still contribute to 
the baryon budget.

\subsection{The Warm-Hot IGM}

The search for the WHIM gas has now begun, using sensitive UV resonance
absorption lines:  primarily the O~VI lithium-like ($2s-2p$) doublet 
(1031.926, 1037.617~\AA) but also the Li-like doublet of 
Ne~VIII (770.409, 780.324~\AA) -- see Savage \etal (2004). The O~VI 
lines can be observed with HST/STIS for IGM absorbers with $z \geq 0.11$,  
while the FUSE spectrographs can observe O~VI essentially down to $z = 0$. 
The K-shell X-ray absorption lines of O~VII (21.602~\AA), O~VIII (18.97~\AA), 
N~VII (24.782~\AA), Ne~IX (13.447~\AA), and a few other highly ionized 
species can (barely) be observed with {\it Chandra} and {\it XMM/Newton}, 
primarily toward X-ray flaring blazars. As mentioned in \S~2, there now  
exists a substantial number (35 to our knowledge) of STIS medium-resolution 
echelle spectra of QSOs. However, fewer than half of these targets have 
sufficient redshift ($z \geq 0.11$) and continuum signal-to-noise ratio 
(S/N $\geq 10$ per resolution element) to facilitate a sensitive search 
for O~VI absorption.  Thus far, only a modest pathlength has been searched 
for O~VI absorbers with HST (Tripp, Savage, \& Jenkins 2000; Savage \etal 
2002). These spectra yield measured baryon fractions of 5--10\% 
(assuming 10\% O/H metallicity), consistent with expectations from 
simulations (Gnedin \& Ostriker 1997; Cen \etal 2001; Fang \& Bryan 2001). 
We caution that these estimates involve large uncertainties, including untested
assumptions of cloud sizes and shapes, metallicities, ionization equilibrium,
and multi-phase structure.   

The multiphase nature of IGM absorbers leads directly to the
classic census problem of ``double counting". In the IGM context, 
this means that the photoionized H~I and O~VI absorbers could count
the same baryons twice.  Recent surveys 
down to 50~m\AA\ equivalent width give H~I line densities approximately
ten times those of the O~VI absorbers:   
$d{\cal N}_{\rm HI}/dz \approx 112 \pm 9$ (Paper IV) and 
$d{\cal N}_{\rm OVI}/dz \approx 14^{+9}_{-6}$ (Savage \etal 2002).  
Many IGM absorbers (Shull \etal 2003) as well as Galactic high velocity 
clouds (Collins \etal 2004) clearly contain gas at vastly different 
temperatures, in order to explain the observed range of ionization stages.  
Deriving the metallicities of such complex systems requires subtle
ionization modeling:  collisional vs.\ photoionization, time-dependent
ionization, range of temperatures, inhomogeneities in density and velocity.   

Theoretical considerations of the shocked IGM (Dav\'e \etal 1999) 
suggest that collisional ionization dominates the stronger O~VI systems, 
although some weak O~VI absorbers 
have been modeled successfully using photoionization equilibrium 
(Tripp \etal 2001). The problem with these photoionization models for
high ions is that they require a large photoionization parameter 
($U \propto J_0/n_H$) to produce sufficient O~VI.  
At fixed ionizing radiation intensity, $J_0$, this requires very
low hydrogen densities, $n_H \approx (10^{-5}$ cm$^{-3}) n_{-5}$.  As
a result, photoionization models with high-$U$ and low-$n_H$ are 
physically implausible, since they produce a low neutral fraction, 
\begin{equation}
   f_{\rm HI} \equiv \frac {n_{\rm HI}} {n_H} \approx 
    (3 \times 10^{-5}) \; n_{-5} \; T_4^{-0.845} \, \Gamma_{-13}^{-1} \; , 
\end{equation}
and unrealistically large absorption pathlengths, 
\begin{equation}
    L_{\rm abs} \equiv \frac {N_{\rm HI}}{n_H \, f_{\rm HI}} 
       \approx (350~{\rm kpc}) \; \Gamma_{-13} \; T_4^{0.845} \; 
        n_{-5}^{-2} \; N_{14.5} \;  .      
\end{equation}
Here, we have scaled the low-redshift parameters to H~I photoionization 
rate $\Gamma_H = (10^{-13}~{\rm s}^{-1}) \Gamma_{-13}$, a temperature 
$(10^4~{\rm K})T_4$, and a typical strong H~I absorber with 
\NHI~$= (10^{14.5}~{\rm cm}^{-2}) N_{14.5}$. Shull \etal (1999) find 
that $\Gamma_{-13} \approx$ 0.3--0.7 at $z \approx 0$; the ionization rate 
can also be written $\Gamma_H \approx (2.5 \times 10^{-14}~{\rm s}^{-1}) I_{-23} 
[4.8/(\alpha_s+3)]$, where $I_{-23}$ is the specific intensity at 1 ryd 
and $\alpha_s \approx 1.8$ is the mean spectral index of the ionizing radiation.     
Thus, some of the photoionized baryons in the O~VI absorbers may have 
been accounted for by the H~I calculation 
in the previous section. However, the hotter, collisionally ionized
O~VI should represent a distinct thermal phase from the H~I.   

A substantial survey of O~VI absorbers may reveal systematics among the 
O~VI absorbers that will allow us to separate the collisionally-ionized 
absorbers from the photoionized absorbers and thus unravel the double-counting 
problem. Ongoing STIS echelle spectroscopy by Tripp, Savage, Howk, and
others will go a long way towards answering these questions and better 
quantifying the ``O~VI forest'' statistics. Equally important to this effort 
is ongoing FUSE spectroscopy of lower-redshift O~VI with complementary \lya\ 
data from {\it Hubble}. To date, we have found 114 \lya\ absorbers
($W_{\lambda} \geq 80$~m\AA) with acceptable O~VI data, and roughly half 
(57 systems) are detected in O~VI at $\geq3 \sigma$ significance (Danforth 
\& Shull 2004). This combined HST/FUSE database can address important 
questions concerning the extent to which metals are spread away from galaxies 
into the IGM. While much data exist and more is scheduled to be obtained, 
these are still early days.

If the study of the O~VI absorbers is still in its infancy, the detection of 
the remainder of the WHIM gas is not yet ``out of the womb''. Simulations 
predict that this hotter WHIM (10$^{6-7}$~K) contains $\geq$ 20--40\% of the 
total local baryons. Nicastro (2003) claims that WHIM gas at 
$T \approx 10^{5.8}$~K has been detected at $z=0$ along several sightlines, 
through absorption lines of O~VI with FUSE and of C~VI, O~VII, O~VIII, 
N~VII, and Ne~IX with {\it Chandra} and/or {\it XMM/Newton}. However, it is 
possible that this gas is not intergalactic at all, but instead resides in 
smaller physical regions in the halo of the Milky Way  (Sembach 2002; Shull 
2003; Fox \etal 2004; Collins \etal 2004). While two groups have claimed WHIM 
detections at $z>0$ in the sightlines of PKS~2155-304 (Fang \etal 2002) 
and H1821+643 with {\it Chandra} (Mathur \etal 2003), neither of these 
results has been confirmed. The PKS~2155-304 detection has been 
refuted by {\it XMM/Newton} observations (Rasmussen \etal 2003; 
Cagnoni \etal 2004). 

Recently, two new WHIM detections were obtained (Nicastro \etal 2004) toward 
the blazar Mrk~421 at $z = 0.03$, while it flared in X-rays. The claimed 
O~VII absorbers lie at $z = 0.011$ (where a \lya\ absorber was identified by 
Shull, Stocke, \& Penton 1996) and $z = 0.027$ (where we see no \lya\ or 
O~VI absorbers to $4\sigma$ limits of 30~m\AA\ and 20~m\AA, respectively). 
The $z = 0.027$ absorber was detected in three X-ray lines (N~VI, N~VII, O~VII)
at $\sim 3 \sigma$ significance each, while the $z = 0.011$ absorber was seen 
in only O~VII at $3.8\sigma$ significance (but not in O~VIII).   
The O~VIII/O~VII is consistent with WHIM at temperatures 
log $T \approx$ 5.7--6.5.  Over the small observed pathlength, 
$\Delta z \approx 0.03$, these $N = 1-2$ detections yield an O~VII line 
density, $N/(\Delta z)$, of 2--4 times that of O~VI, with large  
uncertainty.  The O~VII columns ($\sim 10^{15}$~cm$^{-2}$)
are $\sim 10$ times those typically seen in O~VI.  Therefore, the baryon 
fraction in the O~VII-bearing gas could be anywhere from 20--70\%,
with enormous uncertainties arising from cosmic variance of this one
short sightline.  

A possible one-line WHIM detection in the {\it Chandra} spectrum 
of 3C~120 (McKernan \etal 2003) yields a similar O~VII line density and 
baryon fraction. Obviously, the statistics of these studies are poor,
and the model assumptions substantial.  Sadly, the prospects of further 
observations with {\it Chandra} are remote; when Mrk 421 was observed it 
was the brightest extragalactic source in the sky (the brightest 
extragalactic soft X-ray sources tend to be blazars). Because the velocity 
resolution of {\it Chandra's} Low Energy Transmission Grating is only 
$\sim$ 1000 \kms, one worries about the proximity (900 \kms) of the $z = 0.027$ 
absorber to the blazar redshift ($z = 0.03$).  Thus, the $z = 0.027$ 
absorber might be outflowing gas along a jet and not intervening WHIM at all.  
Taking this observational program to the level of FUSE (or even HST) may
have to await the {\it Constellation-X} grating spectrographs.

\section{The Relationship Between \lya\ Absorbers and Galaxies}

While the baryon census and its evolution are ample reasons for studying the 
local \lya\ forest in detail, it is also only at low redshift that the locations
of \lya\ absorbers and galaxies can be compared accurately. This allows
the relationship between these ``clouds'' and galaxies to be determined to 
some degree of certainty. However, the degree to which absorbers correlate with 
individual galaxies has been controversial. Lanzetta \etal (1995) and Chen
\etal (1998, 2001) argue that the \lya\ absorbers are the very extended halos 
of individual galaxies, while others (Morris \etal 1993; Stocke \etal 1995; 
Impey, Petry, \& Flint 1999; Dav\'e \etal 1999; Papers III and V) argue that 
the absorbers are related to galaxies only through their common association
with large-scale gaseous filaments, arising from overdensities in the
high-redshift universe. Much of the difference between these results primarily 
reflects the column density range of the \lya\ absorbers studied in each case; 
at some level both sides to this argument are correct (see conference
papers in Mulchaey \& Stocke 2002). In this section, we review the current
evidence for the relationship between galaxies and \lya\ absorbers as a function 
of H~I column density.

\subsection{The Damped \lya\ Absorbers (DLAs)}

At the highest column densities (\NHI $\geq10^{20.3}$~\cm2) \lya\ exhibits 
broad damping wings that allow an accurate determination of H~I column density. 
This has proved valuable for studying the evolution of the chemical 
abundance of galaxies with cosmic time, since even quite weak metal lines are 
detectable in these systems with high-quality spectra (e.g., Pettini \etal 1999;
Prochaska \& Wolfe 2000, 2002). DLAs have also been used to measure the evolution 
of the amount of neutral gas in galaxies (Boissier \etal 2003). However, clues 
about the types of galaxies probed by the DLAs are largely circumstantial at 
high redshift. Thick disks of massive galaxies are suggested by the kinematics 
(Wolfe \& Prochaska 2000), but are hardly demanded by the evidence. Surveys of 
DLAs with HST (e.g., Rao \& Turnshek 2000; Turnshek \etal 2002) and their
subsequent imaging with large-aperture ground-based telescopes (Nestor \etal 
2002; Rao \etal 2003) have shown convincingly that no single galaxy population 
dominates the DLAs. Surprisingly, there is a large contribution to DLAs made by 
dwarf and low surface brightness (LSB) galaxies (see Bowen, Tripp, \& Jenkins 
2001 for an 
excellent example) in contrast to the Wolfe \& Prochaska (2000) kinematic analysis. 
Until recently, an apparent contradiction existed between the cross-sectional
areas on the sky covered by high column density H~I at $z=0$ as measured either 
by DLAs or by 21~cm emission maps (Rao \& Briggs 1993). However, new ``blind'' 
21~cm surveys (Zwaan \etal 2001; Rosenberg \& Schneider 2003) show that the 
21-cm cross-sectional area now agrees with the DLA measurement at low-$z$.

These new surveys also show that the galaxy populations detected by these two 
different methods now agree. Rosenberg \& Schneider (2003) find that DLA 
galaxies span a wide range of total H~I masses (two-thirds are between 
$M_{\rm HI} = 10^{8.5-9.7}~M_{\odot}$) and luminosities 0.01--1 $L^*$.
Thus, the chemical abundance data from DLAs at high-$z$ must be interpreted 
with these new HST $+$ ground-based results in mind; that is, the chemical 
abundance evolution refers to H~I disks and amorphous H~I clumps in a wide 
variety of galaxies. This certainly explains the broad spread in DLA 
metallicities at any one redshift.

\subsection{Lyman Limit Absorbers}

In the range \NHI $= 10^{17.3-20.3}$~\cm2, the Lyman Limit System (LLS) absorbers 
constitute a substantial number ($dN/dz \approx$ 1--3 per unit redshift) 
of intervening systems in the spectra of high-$z$ QSOs. As shown by Steidel 
(1995, 1998), strong ($W_{\lambda} \geq 0.3$~\AA) Mg~II absorbers at $z =0.2-1$ 
are invariably also LLSs, although a few DLAs are also present in such samples. 
Physically, this can be understood because H~I and Mg~II have similar ionization 
potentials, so that once hydrogen becomes optically thin, Mg~II ionizes rapidly 
to Mg~III. In an important pre-HST result, Steidel (1995, 1998) found that, 
of 58 LLSs in his sample, 55 could be identified with nearby 
($\leq 50 h^{-1}_{70}$ kpc offsets), bright ($L > 0.1-0.3~L^*$) galaxies at the 
same redshift (see also Bergeron \& Boiss\'e 1991).  These results have been 
strengthened by obtaining ground-based spectra of the underlying stellar 
kinematics of the associated galaxies and HST imaging and UV spectra of the 
C~IV absorptions in these systems (Churchill \etal 2000; Steidel \etal 2002). 
These results suggest that the LLSs are almost exclusively the bound gaseous 
halos of luminous galaxies, and that these halos share the kinematics of the 
underlying stellar disk. 

Recently, HST observations of the quasar 3C~232 probed the halo of the 
modest starburst galaxy NGC 3067.  The spectra show a complex LLS 
with three absorption components at small radial velocity differences 
($\Delta v = -260$, $-130$ and $+170$ \kms) from the nucleus of  NGC 3067,
in ions ranging from Na~I and Ca~II detected optically (Stocke \etal 1991) 
to Mg~I, Mg~II, Fe~II and Mn~II in the near-UV (Tumlinson \etal 1999) to 
Si~IV and C~IV in the far-UV. Surprisingly, all these species share the same 
three velocities (Keeney et al. 2005), which are gravitationally bound to 
NGC 3067; two of the systems, including the strongest at \NHI $=10^{20.0}$
\cm2, are infalling. The 3C~232 sightline has an offset of $8h^{-1}_{70}$ kpc
from the nucleus of NGC 3067; the sightline clearly penetrates extra-planar 
H~I, seen in 21~cm emission with the VLA (Carilli \& van Gorkom 1992). 
The H~I column densities, kinematics, metallicity, spin temperature, and 
inferred size of these clouds are similar to high velocity clouds in the 
Milky Way. Thus, despite a modest starburst ($\geq 0.6~M_{\odot}$~yr$^{-1}$) 
in this $0.5 L^*$ galaxy, even the high ionization gas does not appear to be 
escaping.  In summary, both the detailed HST and ground-based observations of 
this nearby LLS and the cumulative data from the Steidel LLS sample 
are consistent with LLSs being the bound, gaseous halos of luminous galaxies.

\subsection{Weak Metal-line Systems}

Both the HST QSO Absorption Line Key Project and our GHRS/STIS \lya\ survey 
of Papers I--IV have found large numbers of intermediate column density 
(\NHI $= 10^{13-16}$~\cm2) \lya\ absorbers at low redshift. Available 
HST or FUSE spectra are not sensitive enough to detect metals 
(C~II/III/IV, Si~II/III/IV, O~VI) 
in these absorbers, except in a few cases at \NHI $= 10^{15-16}$~cm$^{-2}$. 
For example, metallicities have been measured to 1--10\% of solar values
in strong \lya\ absorbers in the sightlines toward 3C~273 (Sembach \etal 2001;
Tripp \etal 2002), PKS~2155-304 (Shull \etal 1998, 2003), PG~1259+593
(Richter \etal 2004), and PG~1211+143 (Tumlinson \etal 2004).  In an ongoing 
survey, Danforth \& Shull (2004) have found that $\sim$50\% of all \lya\ 
absorbers contain both H~I and O~VI in the \NHI\ range for which high-quality 
FUSE spectra exist.  The ``multiphase ratio", N(H~I)/N(O~VI), varies 
substantially, from $\sim0.1$ to greater than 100 (Shull 2003; Danforth \&
Shull 2004), probably indicating a wide range in shock velocities, and 
possibly O/H abundances, in these multiphase systems.

Given these statistics, it is conceivable that a large fraction of 
$z \approx 0$ \lya\ absorbers at \NHI $= 10^{13-17}$~\cm2 contain
metals. The O~VI non-detections are either at lower ionization (lower
shock velocity) than 
allows significant columns of O~VI, or at slightly lower metallicity than 
the FUSE sensitivity limits can detect. Since most or all \lya\ absorbers 
at $z =2 - 4$ in this column density range contain C~IV and/or O~VI 
absorption lines at metallicities $[Z] = -1$ to $-3$ (e.g., Schaye \etal 
2000), it is not surprising that many low column density, low-$z$ absorbers 
would contain some metals. Thus, some relationship with nearby galaxies 
is expected.  Figure 4 shows the galaxies surrounding the sightline to 
PKS~2155-304, coincident with a cluster of strong \lya\ absorbers,
in which C~IV or Si~III have not been detected to a limit of 
$\sim$3\% solar metallicity (there are two weak O~VI absorbers detected
by FUSE).

\begin{figure}[t]
\epsscale{0.7}
\plotone{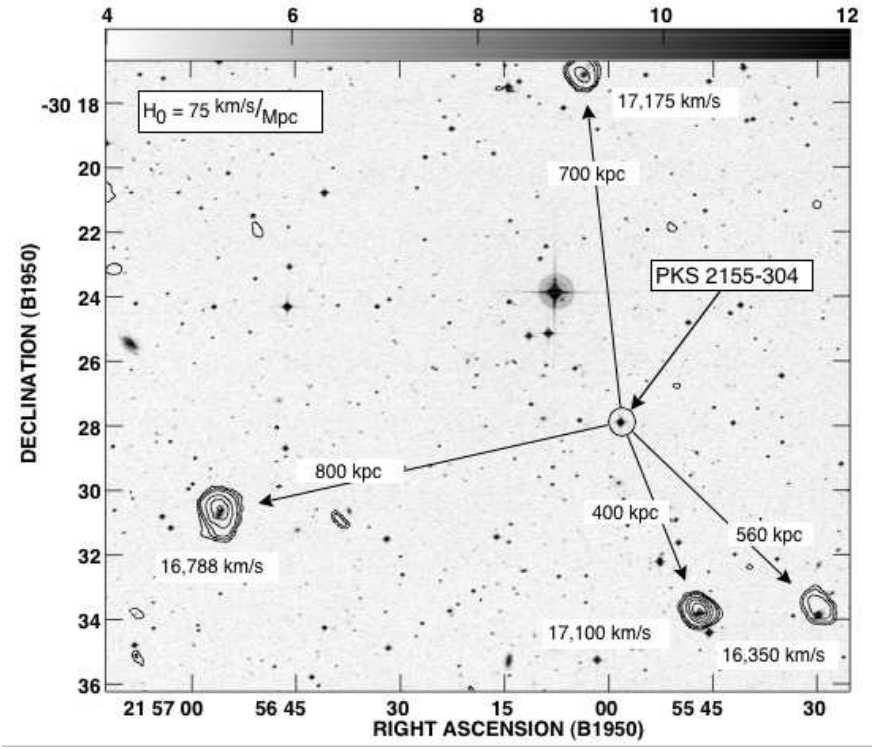}
\caption{\small Optical and VLA field toward PKS~2155-304 (Shull \etal 1998, 
  2003) showing galaxies located at velocities (16,350--17,175 km~s$^{-1}$)
  similar to a cluster of 7 Ly$\alpha$ absorbers.
  These H~I galaxies are located at projected offsets of
  $(400-800)h_{75}^{-1}$~kpc, and the group has an overdensity
  $\delta \approx 10^2$. The multiphase gaseous medium has been detected
  in H~I (Ly$\alpha$, Ly$\beta$, Ly$\gamma$), O~VI, and perhaps O~VIII
  (the claimed X-ray detection by Fang \etal 2002 has not been confirmed). }
\end{figure}

Lanzetta \etal (1995) and Chen \etal (1998, 2001) obtained deep images and 
multi-object spectroscopy for a large number of galaxies in the fields of 
Key Project absorbers. Lanzetta \etal (2002) summarizes the evidence
for a $\sim$30\% success rate in matching redshifts of absorbers and
luminous galaxies within $240 h^{-1}_{70}$~kpc of the sightline.  They then
extrapolate to the conclusion that all FOS-discovered \lya\ absorbers can be 
associated with very 
extended galaxy halos. Chen \etal (1998) found that, with the exception of 
a weak dependence on galaxy luminosity, there is no nearest-galaxy property 
that correlates with absorber properties. On the other hand, using \lya\ 
absorbers at somewhat lower column densities, Morris \etal (1993), Tripp, 
Lu, \& Savage (1998), Impey, Petry, \& Flint (1999) and our Papers III and V 
found no convincing statistical evidence that these GHRS- and STIS-discovered 
absorbers are associated with individual galaxies down to $0.1L^*$ 
luminosities. However, over 3/4 of these weak absorbers are found in galaxy 
large-scale structure filaments (Paper III), with bright galaxies several 
hundred kpc away. Stocke (2002) summarizes the evidence in favor of this 
position. At the extremes of the column density range are obvious examples 
that both of these positions are correct for some absorbers; at the 
high-\NHI\ end of this range, LLSs are so close to the nearby bright galaxy 
that an association of some sort seems inevitable. 

On the other hand, an increasing percentage of lower column density absorbers 
are found in galaxy voids (Papers III and V), more than $3h^{-1}_{70}$ Mpc from 
the nearest bright or faint galaxy (McLin \etal 2001). At intermediate \NHI, 
the absorbers adjacent to gas-rich galaxies do not have velocities
consistent with rotation curves of the nearby galaxy.  This
suggests no kinematic link to the nearby galaxy (C\^ot\'e \etal 2002; 
Putman \etal 2004), in contrast to the Steidel \etal (2002) kinematic analysis 
of LLSs and their associated galaxies that often shows a relationship
between absorber metal-line velocities and the kinematics of the underlying 
stellar disk. 

Recently, the discovery (Stocke \etal 2004) 
of a dwarf ($M_B =-14$, $L = 0.004L^*$) post-starburst 
galaxy $70h^{-1}_{70}$~kpc from a weak-metal line absorber (Figure 5) with 
\NHI $\approx 10^{16}$~\cm2 suggests a solution compatible with all the data 
presented above. Not only do the absorber and galaxy redshifts match to within 
their combined errors, but the absorber metallicity (6\% solar) and the 
mean stellar metallicity of the galaxy ($\sim$10\% solar) approximately match. 
Further, the absorber has an overabundance of silicon to carbon indicative of 
recent supernova type II enrichment.  The galaxy is a pure disk system whose 
optical spectrum has both strong Balmer and metal absorption lines, and no 
evidence for dust or gas (no emission lines and 
$M_{\rm HI} \leq 5 \times 10^6~ M_{\odot}$; van Gorkom \etal 1993). 
From ratios of Lick absorption-line indices, we estimate that the mean stellar 
age in this galaxy is $3.5 \pm 1.5$ Gyrs. Taken together, this information 
provides a consistent picture in which a massive ($\geq 0.3~M_{\odot}$ 
yr$^{-1}$) starburst $\sim 3$ Gyrs ago created enough supernovae to blow away 
the remaining gas from this galaxy into the IGM. Because
the dwarf is quite low mass, this wind can easily escape from the galaxy and 
move to $\sim 100h^{-1}_{70}$ kpc at the required 20--30 \kms\ to create the
metal-line absorber that we see between us and 3C~273.

\begin{figure}[t] 
\epsscale{0.5}
\plotone{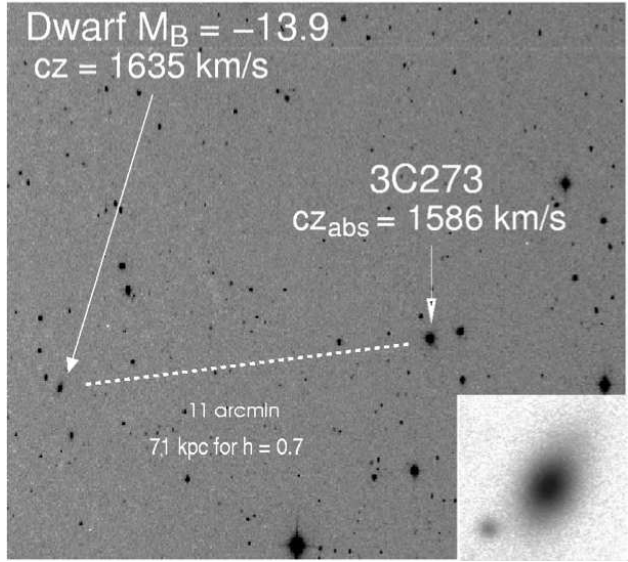}
\caption{\small
Digitized Sky Survey image of the region around 3C~273 (north up, east to 
left) showing the dwarf galaxy located $\sim 70 h^{-1}_{70}$ kpc away from 
the absorber on the sky and at the same recession velocity to within the 
errors ($\pm 50$ \kms) -- see Stocke \etal 2004. 
The R-band image of the dwarf galaxy taken with 
the ARC 3.5m telescope at Apache Point is shown as an insert at lower right. 
The surface brightness profile of this galaxy is well-fitted by a pure 
exponential disk.}
\end{figure}

Because this is the nearest weak metal-line system known, many other similar
absorbers could arise from starburst winds produced by dwarf galaxies too faint 
to be detected by the above surveys (at $z \approx 0.1$, a galaxy like this 
dwarf would have $m_B \sim 24$). This explains why Lanzetta \etal (1995) found 
bright galaxies $(100-300)h^{-1}_{70}$ kpc away that do not have properties that 
correlate with absorber properties. These absorbers were produced not by 
the bright galaxy, but by a dwarf or dwarfs that accompany the bright galaxy. 
This also explains the correlation between absorber locations and 
large-scale structure in the samples of Paper III and Impey,
Petry \& Flint (1999). Because dwarf galaxies are so numerous 
(1--3 Mpc$^{-3}$), if each dwarf galaxy had at least one massive starburst 
that ejected most or all of the gas from the galaxy to a distance of 
$\sim 100h^{-1}_{70}$ kpc, this process would be sufficient to create several 
hundred weak metal-line absorbers per unit redshift.  This number is comparable 
to the line density of all metal-line absorbers at $z=2$.
While this dwarf galaxy would have been much more luminous 
($M_B \approx -16.5$) when it was ``starbursting'', the present-day absence 
of gas means that it will no longer form stars and will eventually fade to 
the approximate luminosity of a Local Group dwarf spheroidal. Thus, the dwarf 
in Figure 5 is the expected intermediate stage
between the ``faint blue galaxies'' seen at $z =0.5-1$ and the present-day dwarf
spheroids (Babul \& Rees 1992). That such ``Cheshire Cat'' galaxies might
be responsible for a large portion of the \lya\ forest was suggested by Salpeter
(1993) and Charlton (1995).

However attractive this solution appears, based upon this one example, an 
equally plausible solution consistent with all available data is that 
large-scale filaments of galaxies are enriched with metals throughout the 
filament to approximately the same metallicity. In this picture, 
metal enrichment is not due to any one galaxy. Indeed, in some 
environments, this almost has to be the case. In the Virgo Cluster, studied
through QSO absorption line probes by Impey, Petry \& Flint (1999), no nearby 
dwarf galaxies were found, despite a galaxy redshift survey complete to 
$M_B = -16$.  In some cases (the $cz=1015$ \kms\ H~I $+$ O~VI absorber 
toward 3C~273, Tripp \etal 2002; and the 1666 \kms\ metal-line absorber 
toward RXJ 1230.8+0115, Rosenberg \etal 2003) weak metal-line absorbers 
have no nearby ($\leq 100h^{-1}_{70}$ kpc) galaxies, despite a considerable 
redshift survey effort.  While our galaxy surveying continues in these 
directions, these two hypotheses may not be easy to discriminate. However, a 
statistical study of the distances that metal-enriched gas extends away from 
galaxies may help to select the better model.

\subsection{\lya-only Absorbers}

Many of the lowest-\NHI\ absorbers could be metal-free, a hypothesis 
consistent with all available data at high and low redshift.
Although HST does not have the sensitivity to address 
this question directly, there is considerable circumstantial evidence in favor
of this hypothesis. While $\sim$80\% of all low column density \lya\ absorbers 
are found in galaxy filaments, the remaining 20\% and an 
increasing percentage of absorbers with decreasing column densities, are found 
$>3 h^{-1}_{70}$ Mpc from the nearest bright galaxy. A deep survey of the few
nearest examples of ``void absorbers''  (McLin et al. 2002) found no faint 
galaxies ($M_B \leq -12$ to $-13.5$, depending on sightline) within several 
hundred kpc of these absorbers.  Thus, there is no plausible 
nearby source for metals in these
absorbers. In Paper IV, we presented a two-point correlation function (TPCF) 
in absorber velocity separation that showed excess power at 
$\Delta v \leq 250$ \kms\ but only for the higher column density absorbers 
in our sample ($W_{\lambda} \geq 65$~m\AA). The lower column density absorbers 
(\lNHI $\leq13$) appear to be randomly distributed in space with no excess
power at any  $\Delta v$ (Paper IV). Searches for metals in the highest 
column density ``void absorbers'' are capable of setting metallicity limits at 
$[Z] \approx -2$ (B. Keeney 2004, private communication), but some high-$z$ 
absorbers contain metals at even lower metallicities. Therefore, even deeper 
HST observations of very bright targets are required to test this interesting 
hypothesis.

\section{The Future}

When HST Servicing Mission 4 was still scheduled, the installation of the 
{\it Cosmic Origins Spectrograph} (COS) on HST was eagerly awaited by 
astronomers who study the local IGM. With 10--20 times the far-UV throughput 
of STIS at comparable (15 \kms) resolution, COS would revolutionize 
low-$z$ \lya\ absorption work by allowing us to obtain high-resolution,
high signal-to-noise ratio spectra of 16$^{\rm th}$ -- $17^{\rm th}$ 
magnitude QSOs in a few orbits.  STIS studies of the IGM require 
ever-increasing numbers of orbits to address new science goals. 
Below is a sampling of the questions that either STIS or COS can 
address; those that truly require COS are noted with asterisks (*):

\begin{enumerate} 

\item {\bf Baryons in the WHIM.} How many are there? 
  Where are they relative to galaxies? What are the ionization processes
  that create the observed absorption-line signatures?

\item {\bf Galaxy halo sizes, shapes, metallicities and kinematics.}$^{*}$  
   A few of the brightest examples can be attempted with STIS, but to 
   provide gaseous halo properties as a function of galaxy type and 
   luminosity is clearly a COS project. 

\item {\bf Low column density \lya\ absorbers.}$^{*}$ How many baryons 
   do these absorbing clouds contain? Do they contain metals and at what 
   metallicity? This is a particularly interesting question for those 
   absorbers in voids, which are the most obvious locations to search for 
   primordial hydrogen clouds, if they exist. Their metals provide an 
   important ``fossil record'' of very early star formation.

\item {\bf Extent of Metal Transport away from Galaxies.}$^{*}$ 
   How far away from star formation sites do metals spread in the IGM? 
   What types of galaxies are responsible for enriching the IGM in 
   metals? Can details of nucleosynthesis be found in the IGM to 
   understand early star formation history and evolution? 

\item {\bf Accurate H~I Column Densities.} 
   FUSE observations of higher Lyman lines can be used to obtain accurate 
   H~I column densities by curve-of-growth techniques.  These are also 
   essential for accurate metallicities. Therefore, this goal is best 
   carried out while FUSE and HST are both still operational.

\item {\bf Physical properties of IGM absorbers.} The column density 
   distribution, $b$-value distribution (temperature),  
   detailed relationship to galaxies and other properties are
   still affected by cosmic variance. Obtaining 10--20 high 
   signal-to-noise STIS medium-resolution echelle spectra or a similar 
   number of high-quality COS spectra of $z \approx 0.5$ QSOs would be 
   sufficient to address concerns about cosmic variance for \lya\ absorbers, 
   but not for the rarer O~VI absorbers. For O~VI, FUSE spectroscopy of 
   already detected \lya\ absorption lines appears to be a viable route 
   to obtaining a sufficient sample. One could also use
   HST/STIS data on QSO targets at $z \geq 0.11-0.12$ (to shift O~VI 
   $\lambda1031.926$ or $\lambda 1037.617$ into the HST/STIS band).  
   This project also means observing fainter targets for many more orbits.
 
\item {\bf Size of IGM absorbers in galaxy filaments and voids.}$^{*}$ 
   QSO pairs available to perform these observations are too faint to 
   observe adequately with STIS.

\end{enumerate}

As nearly everyone has noticed over the past decade, the IGM
has attracted considerable interest from the fields of cosmology,
galaxy formation, and galactic chemical evolution.  Even at low
redshift, a substantial fraction (30--60\%) of the baryons reside in 
the IGM, where they influence the mass infall and chemical history of gas
in galaxies, including the Milky Way. 
If this ``cosmic web" of multi-phase (photoionized, collisionally
ionized) gas truly represents a significant baryon reservoir
(Shull 2003; Sembach 2003), one of the best scientific legacies
from {\it Hubble} (and FUSE) will be an accurate characterization
of its features.  \\

\noindent
{\bf Acknowledgements}  

Financial support for the local Ly$\alpha$ forest work at the University 
of Colorado comes from grants provided through HST GO programs 6593, 8182, 
8125, 8571, 9221, 9506, 9520, and 9778 from the FUSE Project (NASA contract
NAS5-32985, grant NAG5-13004), and theoretical grants from NASA/LTSA 
(NAG5-7262) and NSF (AST02-06042).
We thank Ray Weymann for the faint-galaxy redshift survey
work and his inspirational leadership in this field. We thank Mark Giroux 
for ionization modeling and insightful comments.  We thank our other 
collaborators, Brian Keeney, Jessica Rosenberg, Jason Tumlinson, John Hibbard, 
Charles Danforth, and Jacqueline van Gorkom for valuable contributions and 
discussions, and F. Nicastro and B. Keeney for permission to quote their 
results before publication.

\end{document}